\documentclass[11pt]{article}
\pdfoutput=1 % if your are submitting a pdflatex (\emph{i.e.} if you have
\usepackage{jheppub} % for details on the use of the package, please
\usepackage{amssymb,amsmath,amsthm,graphicx}
\usepackage{latexsym}
\usepackage{bm}
\usepackage[]{hyperref}
\usepackage{cleveref}

\def \be {\begin{equation}}
\def \ee {\end{equation}}
\def \bea {\begin{eqnarray}}
\def \eea {\end{eqnarray}}

\pdfoutput=1

\begin{document}

\title{Weak cosmic censorship with $SU(2)$ gauge field and bound on charge-to-mass ratio}
\author[\sharp]{Yan Song, Si-Yuan Cui, and Yong-Qiang Wang\footnote{yqwang@lzu.edu.cn, corresponding author}}

\affiliation{$^{a}$Lanzhou Center for Theoretical Physics, Key Laboratory of Theoretical Physics of Gansu Province,
	School of Physical Science and Technology, Lanzhou University, Lanzhou 730000, People's Republic of China\\
	$^{b}$Institute of Theoretical Physics $\&$ Research Center of Gravitation, Lanzhou University, Lanzhou 730000, People's Republic of China}

\abstract{We numerically construct the stationary solutions of $SU(2)$ Einstein-Yang-Mills theory in four dimensional anti-de Sitter spacetime. When the $t$ component of $SU(2)$ gauge field is taken to the only nonzero component, we construct a class of counterexamples to the weak cosmic censorship conjecture in Einstein-Maxwell theory.
However, including a nonzero $\phi$ component of $SU(2)$ gauge field, we can argue that there is a minimum value $q_W$, when the charge carried by the $\phi$ component is larger than this minimum value, for sufficiently large boundary electric amplitude $a$, the original counterexamples can be removed and cosmic censorship is preserved.}

\maketitle

\section{Introductions}\label{sec1}

Recently, a possible connection between weak cosmic censorship conjecture (WCCC) \cite{Penrose:1969pc,Wald:1974mk,Sorce:2017dst,Yang:2020czk,Qu:2020nac,Feng:2020tyc} and weak gravity conjecture (WGC) \cite{ArkaniHamed:2006dz,Palti:2019pca,Cheung:2014vva} was proposed in \cite{Crisford:2017gsb}. It is the statement that if the weak gravity conjecture holds, a class of counterexamples \cite{Horowitz:2016ezu} to cosmic censorship in Einstein-Maxwell theory with negative cosmological constant can be removed. These counterexamples lead to infinitely large curvatures visible to distant observers, through the infinite growth of the electric field in the bulk, which has been numerically proved in \cite{Crisford:2017zpi}. However the results proposed in \cite{Crisford:2017gsb} show that including a charged scalar field with sufficiently large charge, the curvature will not grow without bound. 
Note that the minimum value of charge-to-mass ratio necessary to save cosmic censorship is precisely the one predicted by the weak gravity conjecture, which states that to be consistent with quantum gravity, any low energy effective theory with gravity coupled to $U(1)$ gauge field must involve a charged particle with charge-to-mass ratio $q/m\geq1$. In addition, this connection between weak gravity and cosmic censorship is also proved to hold when adding a second Maxwell field or a dilaton field  \cite{Horowitz:2019eum}. In our previous works, we also show that taking account into the self-interaction of scalar field \cite{Song:2020onc} or Born-Infeld electrodynamics \cite{Hu:2019rpw}, cosmic censorship can also be saved by assuming the charged scalar field with charge larger than a minimum.
  
In the context of AdS/CFT correspondence \cite{Maldacena:1997re,Gubser:1998bc,Witten:1998qj,Aharony:1999ti}, the model of gravity coupled to $SU(2)$ field is typically used to construct p-wave holographic superconductor \cite{Cai:2015cya,Cai:2013aca,Gubser:2008wv,Basu:2009vv,Cai:2010zm,Akhavan:2010bf}, which is a type of spatial anisotropic holographic superconductors. For the p-wave superconductor model the condensate of charged field (or the condensate of spatial component of charged field) spontaneously breaks the $U(1)$ gauge symmetry, and picks out a special spatial direction, resulting in a spatial anisotropic superconductor. A similar model is the so called s-wave holographic superconductor \cite{Hartnoll:2008vx,Hartnoll:2008kx}, which is a type of spatial isotropic superconductor. The simplest p-wave holographic superconductor was first proposed based on the $SU(2)$ Einstein-Yang-Mills theory in asymptotically anti-de sitter (AdS) spacetime \cite{Gubser:2008wv}. In this model the spatial components of $SU(2)$ field condense and thus the $U(1)$ gauge symmetry, which generated by a $SU(2)$ generator associated to the time component, is spontaneously broken. In addition, another type of p-wave holographic superconductors has also been constructed in Einstein-Maxwell-vector theory in \cite{Cai:2015cya,Cai:2013kaa}.

In this work we generalize the study of the relation of cosmic censorship and the scalar field charge discussed in \cite{Crisford:2017gsb}, and discuss the weak cosmic censorship with $SU(2)$ gauge field.
Motivated by the p-wave holographic superconductor model, we construct the stationary solutions of the $SU(2)$ Einstein-Yang-Mills theory in four dimensions with asymptotically AdS boundary conditions.  It is clear that when the $SU(2)$ gauge field has only a nonzero $t$ component there is a class of counterexamples to cosmic censorship. We will discuss if adding a nonzero spatial component with sufficiently large charge can remove the original counterexamples and save the cosmic censorship, just like the results proposed in the case of scalar field \cite{Crisford:2017gsb}. 

The paper is organised as follows. In Sec.\ref{sec2}, we set up a model of gravity coupled to $SU(2)$ gauge field using the $SU(2)$ Einstein-Yang-Mills theory in four dimensional AdS spacetime. In Sec.\ref{sec3}, we constructed the solutions when the $SU(2)$ gauge field has only a nonzero $t$ component and when the $SU(2)$ gauge field has both nonzero $t$ and $\phi$ component, with $t$ being time coordinate and $\phi$ being spatial coordinate, and study their properties. Finial, we conclude our results in Sec.\ref{sec4}.

\section{The setup}\label{sec2}
\subsection{Action and equations of motion}

In this section, we set up the model in four dimensional $SU(2)$ Einstein-Yang-Mills theory with negative cosmological constant. The bulk action reads
\begin{subequations}
\begin{align}
S=\frac{1}{16\pi G}\int\!d^{4}x\,\sqrt{-g}\,\left[\left(R-\Lambda\right)-\frac{1}{2}\,F_{\mu\nu}^{a}F^{a\mu\nu}\right]\,\,,
 \label{eq:action}
\end{align}
where $\Lambda=-\frac{6}{L^{2}}$ is the cosmological constant in terms of the AdS length scale $L$ and we set $L=1$ and $G$ is the gravitational constant. $F_{\mu\nu}^{a}$ is the $SU(2)$ gauge field strength tensor and we take
\begin{align}
F_{\mu\nu}^{a}=\partial_{\mu}A_{\nu}^{a}-\partial_{\nu}A_{\mu}^{a}+q\epsilon^{abc}A_{\mu}^{b}A_{\nu}^{c}\,.
\end{align}
\label{eqs:action}
\end{subequations}
$A=A^a_\mu\tau^a dx^\mu$ is the $SU(2)$ gauge potential, with $A^a_\mu$ being the $SU(2)$ gauge field matrix-valued elements and $\tau^a$ being $SU(2)$ generators associated with the Pauli matrix $\sigma^a$ by $\tau^a=\sigma^a/2i$. $q$ is the coupling constant for $SU(2)$ gauge field coupled to gravity. In this work we can regard the coupling $q$ as the charge of the component of the $SU(2)$ gauge field. The tensor $\epsilon^{abc}$ is the full anti-symmetric tensor, with $(a,b,c)=1,2,3$. Without loss of generality, we set $\epsilon^{123}=1$.

From the action (\ref{eqs:action}), the equations of motion can be derived as
\begin{subequations}
\begin{align}
\label{eq:Einsteineq}
R_{\mu\nu}+\frac{3}{L^{2}}g_{\mu\nu}=T_{\mu\nu}-\frac{1}{3}T_{\rho}^{\rho}g_{\mu\nu}\,,\qquad T_{\mu\nu}=F^{a}_{\rho\mu}F^{a\rho}_{\nu}-\frac{1}{4}g_{\mu\nu}F^{a}_{\rho\sigma}F^{a\rho\sigma}\,\, 
\end{align}
and
\begin{align}
\label{eq:YangMillseq}
\nabla_{\mu}F^{a\mu\nu}=-q\epsilon^{abc}A_{\mu}^{b}F^{c\mu\nu}\,,
\end{align}
where $T_{\mu\nu}$ is the energy-momentum tensor for Yang-Mills field. 
\label{eqs:EOM}
\end{subequations}

\subsection{The numerical method}
It is known that the analytical solutions of the $SU(2)$ Einstein-Yang-Mills theory is difficult to be obtained. Therefore we instead find the numerical solutions of the coupled equations of motion (\ref{eqs:EOM}) by using the DeTurck method, which was proposed in \cite{Headrick:2009pv} and reviewed more detail in \cite{Dias:2015nua}. Following the procedure of the DeTurck method, we add a gauge-fixing DeTurck term, $\nabla_{(a}\xi_{b)}$, to the original Einstein equation (\ref{eq:Einsteineq}) and deform it into the following solvable elliptic equation, which is the so-called Einstein-DeTurck equation:
\begin{eqnarray}
\label{eq:EinsteinDeTurck}
R_{\mu\nu}+\frac{3}{L^{2}}g_{\mu\nu}-\nabla_{(a}\xi_{b)}=T_{\mu\nu}-\frac{1}{3}T_{\rho}^{\rho}g_{\mu\nu}\,,
\end{eqnarray}
where $\xi^{a}=[\Gamma_{cd}^{a}(g)-\Gamma_{cd}^{a}(\bar{g})]g^{cd}$, and $\Gamma_{cd}^{a}(\bar{g})$ is the Levi-Civitta connection with respect of the reference metric $\bar{g}$.

Note that the general relativity is invariant under coordinate transformation. We have to impose a special gauge choice to fix the residual degree of freedom, to derive the unique stationary solutions for Einstein equation. In the DeTurck method, the reference metric has the same asymptotic structure and gauge choice as that for the desired metric, and thus by giving an appropriate reference metric we can fix the gauge choice for the desired metric.

It is clear that the Einstein-DeTurck equation (\ref{eq:EinsteinDeTurck}) gives rise to a well-defined elliptic problem and when finding static solutions the elliptic equations can be relaxed by using the Newton-like method. In this work we employ the Newton-Raphson method to integrate it iteratively and use the finite element method to fine the grid in the whole intergration region $[0,1]\times[0,1]$.
We choose the grid size as 200$\times$300, and set them as nonuniform grid to refine the important regions for numerical calculation.
In addition, to ensure the results reasonable, we demand the relative error for stationary solutions less than $10^{- 5}$.

\subsection{Ansatzs and boundary conditions}

We restrict ourself to study a static and axisymmetric metric. In order to obtain a simpler metric form, it is necessary to adopt an appropriate coordinate system, such that there are two Killing vectors, $\partial_t$ and $\partial_\phi$, with $t$ being time coordinate and $\phi$ being angular coordinate. We choose the same coordinates as in \cite{Crisford:2017gsb}. To obtain this we start from the metric for the pure AdS spacetime in the standard Poincar\'e coordinates $(t, r, z, \phi)$
\begin{equation}
\mathrm{d}s^2=\frac{L^2}{z^2}\left[-\mathrm{d} t^2+\mathrm{d} r^2+r^2 \mathrm{d} \phi^2+\mathrm{d}  z^2\right]\,,
\label{eq:pureadszr}
\end{equation}
where both coordinates $r$ and $z$ are in range $[0, \infty)$. By taking a coordinate transform 
\begin{eqnarray}\label{eq:trans}
z=\frac{y\sqrt{2-y^{2}}}{1-y^{2}}(1-x^{2})\,,\,\,\,\,\,\,\,\,
r=\frac{y\sqrt{2-y^{2}}}{1-y^{2}}x\sqrt{2-x^{2}}\,,
\end{eqnarray}
the line element in (\ref{eq:pureadszr}) can be rewritten in polar-like coordinates $(x, y)$ as
\begin{equation}
\mathrm{d}s^2=\frac{L^2}{\left(1-x^2\right)^2}\left[-\frac{\left(1-y^2\right)^2 \mathrm{d}  t^2}{y^2 \left(2-y^2\right)}+\frac{4\,d x^2}{2-x^2}+\frac{4\,\mathrm{d} y^2}{y^2 \left(1-y^2\right)^2\left(2-y^2\right)^2}+x^2 \left(2-x^2\right) \mathrm{d} \phi^2\right]\,,
\label{eq:pureadsxy}
\end{equation}
where both coordinates $x$ and $y$ are in range $[0, 1]$. 

As above, the line element  (\ref{eq:pureadsxy}) describes only the simplest pure AdS spacetime. To construct the static solutions with nonzero amplitude $a$ we need to give the most general ansatz. Under the constraint of the symmetries of our solutions, the most general ansatz for static and axisymmetric solutions take the following form
\begin{eqnarray}\label{eq:ansatz}
ds^{2}=\frac{L^{2}}{(1-x^{2})^{2}}[-\frac{(1-y^{2})^{2}\,Q_{1}\,dt^{2}}{y^{2}(2-y^{2})}+\frac{4\,Q_{4}}{2-x^{2}}(dx+\frac{Q_{3}}{1-y^{2}}dy)^{2} \nonumber\\
+\frac{4\,Q_{2}\,dy^{2}}{y^{2}(1-y^{2})^{2}(2-y^{2})^{2}}+x^{2}(2-x^{2})\,Q_{5}\,d\phi^{2}]\,.
\end{eqnarray}
where $Q_i(i=1,2,..,7)$ are the unknown functions, with functions $Q_i$ dependent only on the coordinates $x$ and $y$. To use the DeTurck method, we take the reference metric to be $Q_{1}=Q_{2}=Q_{4}=Q_{5}=1$ and $Q_{3}=0$. And for $SU(2)$ Yang-Mills field we take
\begin{subequations}\label{eqs:ymansatz}
\begin{align}
\label{eq:ansatz2}
A=A^a_\mu\tau^a dx^\mu=\psi(x,y)\,\tau^{3}\,dt+w(x,y)\,\tau^{1}\,d\phi\,,
\end{align}
with
\begin{align}
\label{eq:ansatz3}
\psi(x,y)=&L\,Q_6(x,y)\,,\\
 \label{eq:ansatz4}
w(x,y)=(1-x^{2})&^{\Delta}y^{\Delta}(2-y^{2})^{\frac{\Delta}{2}}\,Q_{7}\,.
\end{align}
\end{subequations}
The $SU(2)$ generator $\tau^{3}$ generates the $U(1)$ subgroup of the $SU(2)$ nonabelian gauge group, while the other $SU(2)$ generator $\tau^{1}$ generates the $\phi$ component of $SU(2)$ field, which is charged under the $U(1)$ gauge symmetry through the nonlinear coupling of the Yang-Mills equations. 
In the context of AdS/CFT correspondence, $\Delta$ is the conformal dimension of the boundary operator $\mathcal{O}$ dual to the $\phi$ component of $SU(2)$ gauge field, playing the role of mass of the $\phi$ component in AdS spacetime.

When $w(x,y)=0$, the $t$ component of $SU(2)$ gauge field is the only nonzero component, and thus the $SU(2)$ Einstein-Yang-Mills theory is identical to the Einstein-Maxwell theory. It is known that for $w(x,y)=0$, there is a branch of solutions of the $SU(2)$ Einstein-Yang-Mills theory in asymptotically AdS spacetime, which describes the Reissner-Nordstr\"om-anti-de Sitter black hole solutions \cite{Basu:2009vv}. However, when $w(x,y)\neq 0$,  the $\phi$ component condenses and the $U(1)$ gauge symmetry is spontaneously broken, resulting in an analogue of p-wave holographic superconductor.

The integration region for Einstein equation is a finite region $[0,1]\times[0,1]$ in $(x,y)$ coordinates.
In the following we impose the boundary conditions on the stationary solutions, by giving the boundary value of the unknown functions $Q_i$ and their derivatives. 

We first consider the case for $w(x,y)=0$. In this case for gravitational field and the $t$ component we choose to impose the same boundary conditions as given in \cite{Horowitz:2016ezu}. It is because when $w(x, y)=0$, the $SU(2)$ Einstein-Yang-Mills theory reduces to the Einstein-Maxwell theory.
At boundary $x=1$, corresponding to conformal boundary, we choose the conformal boundary metric at asymptotic infinity to be flat and the asymptotic form of vector potential to have only a nonzero time component, which imposes
  \begin{eqnarray}\label{eq:bdyinfty}
Q_{1}(1,y)=Q_{2}(1,y)=Q_{4}(1,y)=Q_{5}(1,y)=1, \quad Q_{3}(1,y)=0,\; \\ \nonumber
\qquad\qquad\qquad Q_{6}(1,y)=a\,p(y)=a\,(1-y^{2})^{n}. \qquad\qquad\qquad\qquad 
\end{eqnarray}
As shown above $a$ is the amplitude of the boundary electric field and $p(y)$ is its fall-off profile denoted by a constant $n$. It is shown in \cite{Horowitz:2014gva} that one has to take $n>1$ in order to construct these counterexamples in Einstein-Maxwell theory. At boundary $x=0$, corresponding to the axis of symmetry, we impose
\begin{eqnarray}\label{eq:bdyaxis}
\frac{\partial {Q_{1}(0,y)}}{\partial x}=\frac{\partial {Q_{2}(0,y)}}{\partial x}=\frac{\partial {Q_{4}(0,y)}}{\partial x}=\frac{\partial {Q_{5}(0,y)}}{\partial x}=\frac{\partial {Q_{6}(0,y)}}{\partial x}=0, \;\;Q_{3}(0,y)=0.
\end{eqnarray}
At boundary $y=0$, corresponding to the intersection point of conformal boundary and axis of symmetry, $r=z=0$, we impose
\begin{eqnarray}\label{eq:bdypoint}
Q_{1}(x,0)=Q_{2}(x,0)=Q_{4}(x,0)=Q_{5}(x,0)=1, \;\;\;\;\;Q_{3}(x,0)=0, \;\;\;\; Q_{6}(x,0)=a.
\end{eqnarray}
At boundary $y=1$, corresponding to the Poincar\'e horizon for $n> 1$, we impose
\begin{eqnarray}\label{eq:bdyhorizon}
Q_{1}(x,1)=Q_{2}(x,1)=Q_{4}(x,1)=Q_{5}(x,1)=1, \;\;\;\;  Q_{3}(x,1)=Q_{6}(x,1)=0.
\end{eqnarray}

When $w(x, y)\neq 0$, we need to additionally impose the boundary conditions for the $\phi$ component of $SU(2)$ gauge field.
For $\Delta=0$ the boundary condition is imposed to be
\begin{eqnarray}\label{eq:bdydlt0}
Q_{7}(1,y)=Q_{7}(0,y)=Q_{7}(x,0)=Q_{7}(x,1)=0\,,
\end{eqnarray}
while for $\Delta=1$ the boundary condition is imposed to be
\begin{eqnarray}\label{eq:bdydlt1}
\frac{\partial {Q_{7}(1,y)}}{\partial x}=0\,, \frac{\partial {Q_{7}(x,0)}}{\partial y}=0\,,\;\;\;\; Q_{7}(0,y)=Q_{7}(x,1)=0\,.
\end{eqnarray}
Note that it is convenient to take $\Delta=1$ for the condensate calculation and $\Delta=0$ for the Kretschmann scalar and minimal charge of the $\phi$ component. In this work we will present our results with these values of $\Delta$. For different values of $\Delta$, the conclusions do not change qualitatively.

With the appropriate boundary conditions as well as the ansatz for the metric and Yang-Mills field, we can construct the smooth stationary solutions of the equations of motion (\ref{eqs:EOM}) with $w(x,y)=0$ and with $w(x,y)\neq0$ by using the DeTurck method, and further investigate their properties to study the weak cosmic censorship with $SU(2)$ gauge field.

\section{The results}\label{sec3}
\subsection{Solutions with $w(x,y)=0$}

According to the ansatz of $SU(2)$ Yang-Mills field from Eqs.(\ref{eqs:ymansatz}), when $w(x,y)=0$ the $SU(2)$ gauge field has only a nonzero $t$ component and thus the Einstein-Yang-Mills theory reduces to the Einstein-Maxwell theory. As proposed in \cite{Horowitz:2016ezu}, there is a class of counterexamples to cosmic censorship in Einstein-Maxwell theory. In this subsection we construct the stationary solutions with $w(x,y)=0$ in the Einstein-Yang-Mills theory, and find that these solutions indeed yield regions of infinitely large curvature once beyond a maximum amplitude $a_{max}$, as discussed in \cite{Horowitz:2016ezu}. This indicates the violation of the weak cosmic censorship conjecture.

\begin{figure}[h!]
\begin{center}
\includegraphics[height=.25\textheight,width=.3\textheight, angle =0]{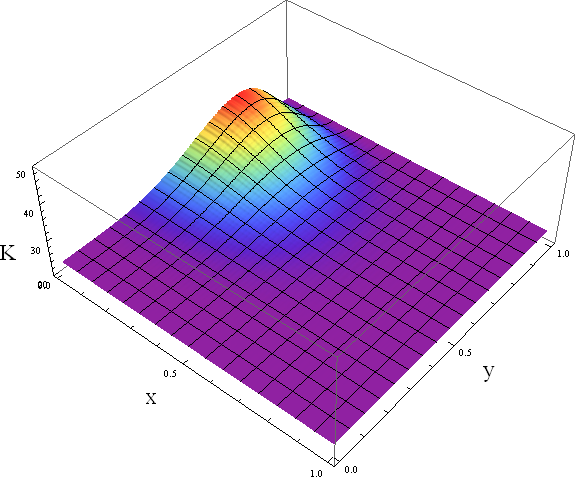}
\includegraphics[width=.35\textheight, angle =0]{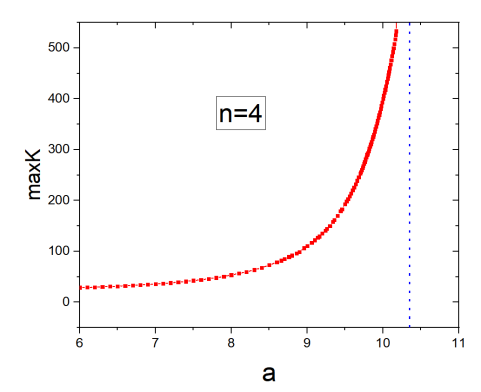}
\end{center}
\caption{\label{fig:kretschmann}$Left$: Distribution of the Kretschmann scalar $K$ over the whole integration domain, for the boundary amplitude $a=7$. The maximum of scalar $K$ locates at axis of symmetry while away from this axis, the $K$ still remains its value for pure AdS, $K=24/L^4$. $Right$: The maximum of the Kretschmann scalar $K$ seems to blow up at a maximum amplitude $a=a_{max}$, marked by the blue vertical dotted line, where a naked singularity appears. For both panels, $n=4$.}
\end{figure}

To describe the growth of the spacetime curvature with increasing the amplitude $a$, we calculate a curvature invariance, the so called Kretschmann scalar \cite{Cherubini:2002gen}$K=R_{\alpha\beta\gamma\delta}R^{\alpha\beta\gamma\delta}$, where $R_{\alpha\beta\gamma\delta}$ is the Riemann curvature tensor. In particular, at $a=0$ the solutions reduce to the pure AdS, for which the Kretschmann scalar is a constant, $K=24/L^4$. Here we display the distribution of Kretschmann scalar $K$ in the whole integral domain for $a=7$ and $n=4$. As shown in the left panel of Fig.\ref{fig:kretschmann}, the maximum value of the Kretschmann scalar $K$ is represented by the red peak while the purple region marks the value near that for pure AdS. It is clear that as amplitude $a$ increases, the maximum value of the scalar $K$ appears at the axis of symmetry, while away from this axis the scalar $K$ remains the value for pure AdS, $K=24/L^4$.
 
When the boundary electric amplitude $a$ continues to increase to a maximum value, the solutions with $w(x,y)=0$ become singular.
This result is shown in the right panel of Fig.\ref{fig:kretschmann}, which displays the maximum value of the Kretschmann scalar $K$ as a function of the amplitude $a$ for $n=4$.
We find that smooth solutions exist only up to a maximum amplitude $a_{max}=10.2$, which is marked by the vertical blue dotted line. Once the amplitude $a$ is slightly larger than this maximum $a_{max}$, the maximum value of the scalar $K$ will grow without bound, indicating the appearance of a naked singularity. The weak cosmic censorship conjecture is thus violated.

From above we argue that when the $SU(2)$ gauge field has only a nonzero $t$ component, we can construct a class of counterexamples to cosmic censorship by increasing the amplitude $a$ to a finite amplitude. Note that these counterexamples we constructed are quantitatively just the same as the ones constructed in  \cite{Horowitz:2016ezu}. In next subsection we will discuss if these counterexamples can be removed by including a nonzero $\phi$ component of $SU(2)$ gauge field with sufficiently large charge, just like the case of a charged scalar field \cite{Crisford:2017gsb}.

\subsection{Solutions with $w(x,y) \neq0$}
\subsubsection{Stability analysis}

We want to check whether the solutions with $w(x, y)=0$ become unstable with the $\phi$ component added and lead to the full nonlinear solutions with $w(x, y)\neq0$.
Using the linear perturbation theory discussed in \cite{Dias:2015nua}, we can firstly consider a perturbatively small $\phi$ component, and analyze the stability of background with $w(x, y)=0$ under the $\phi$ component perturbation \cite{Horowitz:2013jaa,Costa:2017tug}. In this case the back-reaction of the $\phi$ component on the background with $w(x, y)=0$ can be ignored. Therefore, the equation of motion of Yang-Mills field given by Eq.(\ref{eq:YangMillseq}) reduces to a linear perturbation equation about a fixed background with $w(x, y)=0$, which can be considered as a eigenvalue equation for charge $q$. 

\begin{figure}[h!]
\begin{center}
\includegraphics[width=.4\textheight, angle =0]{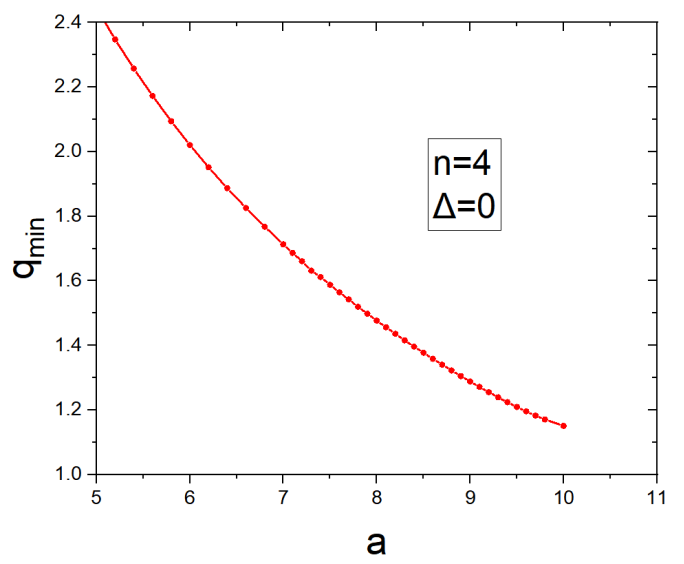}
\end{center}
\caption{\label{fig:stability} The minimal charge $q_{min}$ as a function of the amplitude $a$ for $n=4$ and $\Delta=0$.}
\end{figure}

Solving the linear perturbation equation, we obtain the lowest eigenvalue $q_{min}$ need for a zero mode \cite{Dias:2015nua}. The results is shown in Fig.\ref{fig:stability}, where we plot the minimal charge $q_{min}$ as a function of amplitude $a$ for $n=4$ and $\Delta=0$. The minimal charge $q_{min}$ decreases with $a$ and the zero-mode curve ends at a finite maximum amplitude $a=a_{max}$, corresponding to a lowest value of $q_{min}$, $q_{amax}$. As expected, $a_{max}$ is just the maximum amplitude for smooth solutions with $w(x, y)=0$, where a naked singularity appears. This is because the back-reaction of the $\phi$ component can be neglected when the $\phi$ component is perturbatively small. 

Since zero modes indicate the boundary between stable and unstable solutions, just like the case of a charged scalar field, the solutions with $w(x, y)=0$ are unstable above the curve but stable below it. Therefore, if the $\phi$ component is added with sufficiently large charge, the solutions with $w(x, y)=0$ become unstable and the original counterexamples we constructed to cosmic censorship are no longer valid.
To test whether cosmic censorship can be preserved in the presence of the $\phi$ component, we must take into account the backreaction of the $\phi$ component on the background solutions with $w(x,y)=0$, to construct the full nonlinear solutions with $w(x,y)\neq 0$ and investigate their properties. 

\subsubsection{Full nonlinear solutions}

According to the mechanism studied in \cite{Hartnoll:2008vx}, we can expect that in the presence of the $\phi$ component with sufficiently large charge, the static solutions with $w(x, y)=0$ become unstable to condense the $\phi$ component, forming a new stationary solution with $w(x,y)\neq0$. To measure the size of the $\phi$ component of $SU(2)$ gauge field, we calculate the condensate \cite{Cai:2015cya,Ammon:2009xh} of the $\phi$ component. According to the AdS/CFT correspondence, the condensate of $\phi$ component can be formulated by the expectation value $\langle \mathcal{O}\rangle$, where the operator $\mathcal{O}$ is the boundary operator dual to the $\phi$ component of $SU(2)$ gauge field. In order to calculate the $\phi$ component condensate it is convenient to take $\Delta = 1$.
From the ansatz of the $\phi$ component $w(x, y)$ defined in Eq.(\ref{eq:ansatz4}), the $\phi$ component condensate can be expressed in terms of the unknown function $Q_7$, which reads $\langle \mathcal{O}\rangle=(1-y^2)^{\Delta} Q_7$. Note that the charge carried by the $\phi$ component is required to be large enough such that the condensation occurs. In this work we take $q=1.15>q_{amax}$, where $q_{amax}$ is the minimal charge required to induce instability for smooth solutions with $w(x,y)=0$, corresponding to the maximum amplitude $a_{max}$.

\begin{figure}[h!]
\begin{center}
\includegraphics[height=.25\textheight,width=.3\textheight, angle =0]{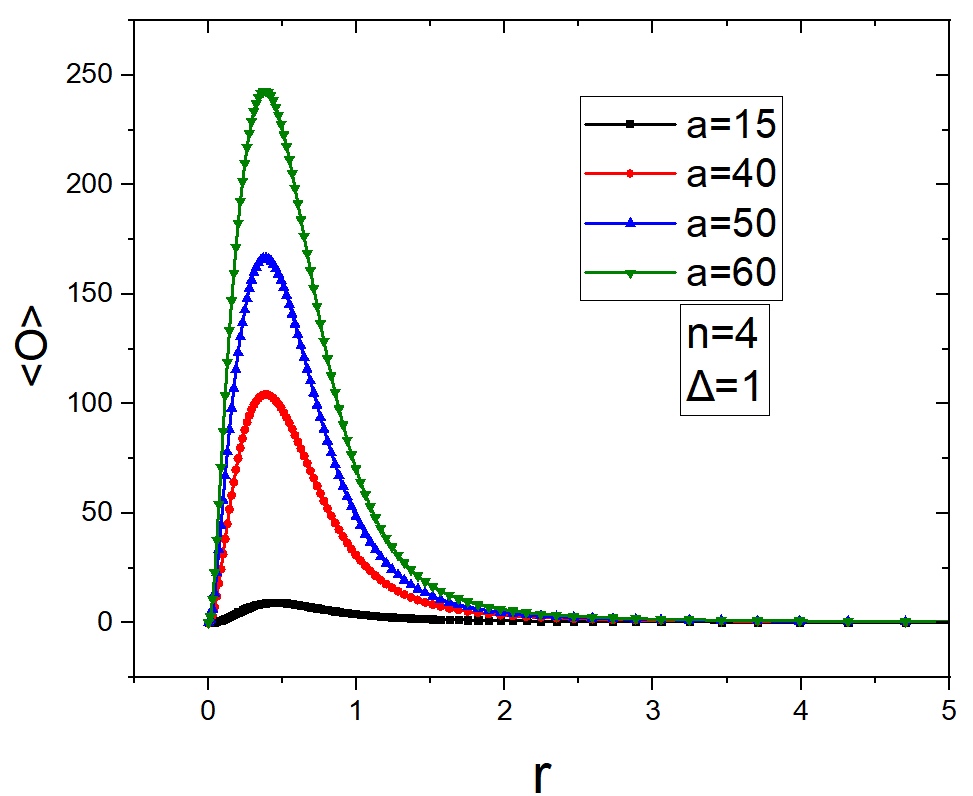}
\includegraphics[height=.25\textheight,width=.3\textheight, angle =0]{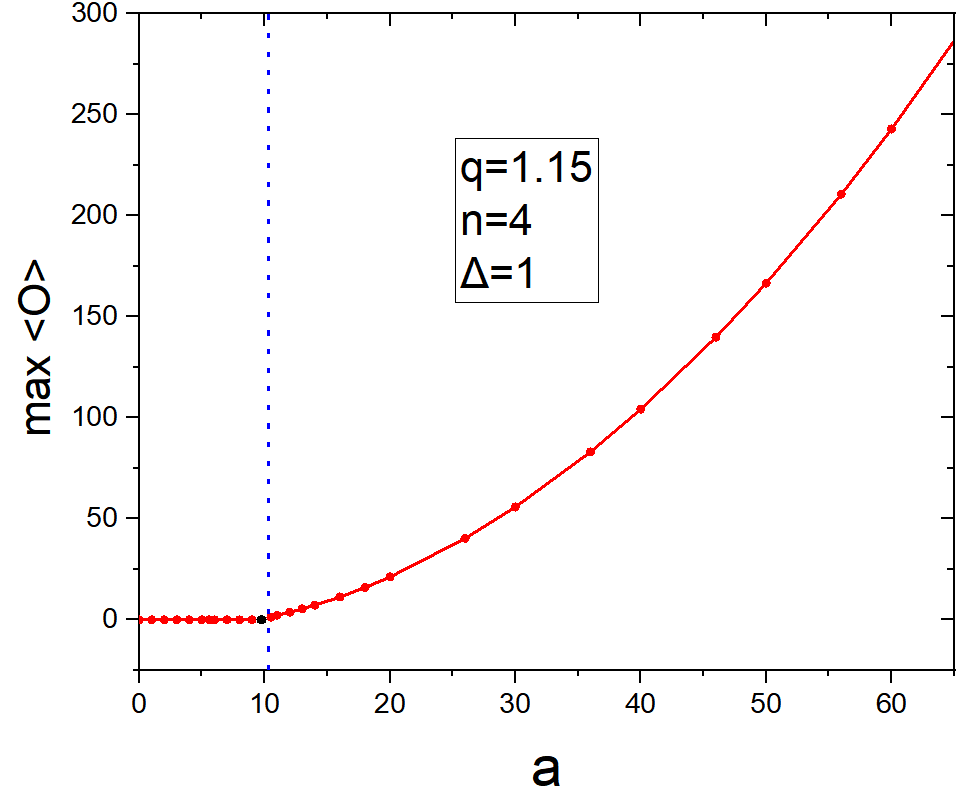}
\end{center}
\caption{ $Left$: Distribution of the condensate of $\phi$ component $\langle \mathcal{O}\rangle$ at conformal boundary. From up to down, the green, blue, red and black solid curves represent the boundary amplitude $a=60, 50, 40, 15$, respectively. $Right$: Maximum of the condensate of $\phi$ component, $max \langle \mathcal{O}\rangle$, as a function of amplitude $a$. The black point indicates the onset for condensate and the vertical dotted blue line marks the maximum amplitude of smooth solutions with $w(x,y)=0$. Both panels are for $n=4$, $\Delta=1$ and $q=1.15$.}
\label{fig:condensate}
\end{figure}

In the left panel of Fig.\ref{fig:condensate}, we present the distribution of the $\phi$ component condensate, $\langle \mathcal{O}\rangle$, at conformal boundary for several values of the boundary amplitude $a$ for fixed $n=4$, $q=1.15$ and $\Delta=1$. The green, blue, red and black solid curves are calculated for $a=60, 50, 40, 15$, respectively. As shown in the plot, the solutions with $w(x,y)\neq0$ indeed exist for all values of amplitude we discussed, even for the amplitudes larger than the maximum amplitude $a_{max}$ when $w(x,y)=0$. For different values of $a$, the condensate of the $\phi$ component always reaches to a maximum value at an almost same $r$, and decays to zero both at origin $r=0$ and at infinity. The maximum value of condensate, $max\langle \mathcal{O}\rangle$, appears to always increase as the boundary amplitude $a$ increases.

In the right panel of Fig.\ref{fig:condensate}, the maximum of condensate $max \langle \mathcal{O}\rangle$ against amplitude $a$ is presented for $q=1.15$, where the black point indicates the onset for condensing the $\phi$ component, $a=10$, and the blue vertical dotted line marks the original maximum amplitude, $a_{max}=10.2$. For the amplitudes larger than the onset $a=10$, the maximum values of condensate $max\langle \mathcal{O}\rangle$ are always nonzero and increase as the boundary amplitude $a$ is increased. This results indicate that as we increase $a$ above the onset value at fixed charge $q=1.15$, the smooth stationary solutions with the nonzero $\phi$ component always exist for large amplitude. It seems that the solutions exist even for arbitrarily large amplitude $a$. 

Thus, we expect that at fixed charge $q=1.15$ there is no maximum amplitude above which the solutions become singular, even for amplitudes much larger than the original maximum amplitude $a_{max}$. It is to say that smooth solutions exist for arbitrarily large amplitude $a$, and the naked singular would not appear. We can no longer violate the weak cosmic censorship conjecture again by slowly increasing the boundary amplitude $a$ to a finite maximum.

\begin{figure}[h!]
\begin{center}
\includegraphics[width=.4\textheight, angle =0]{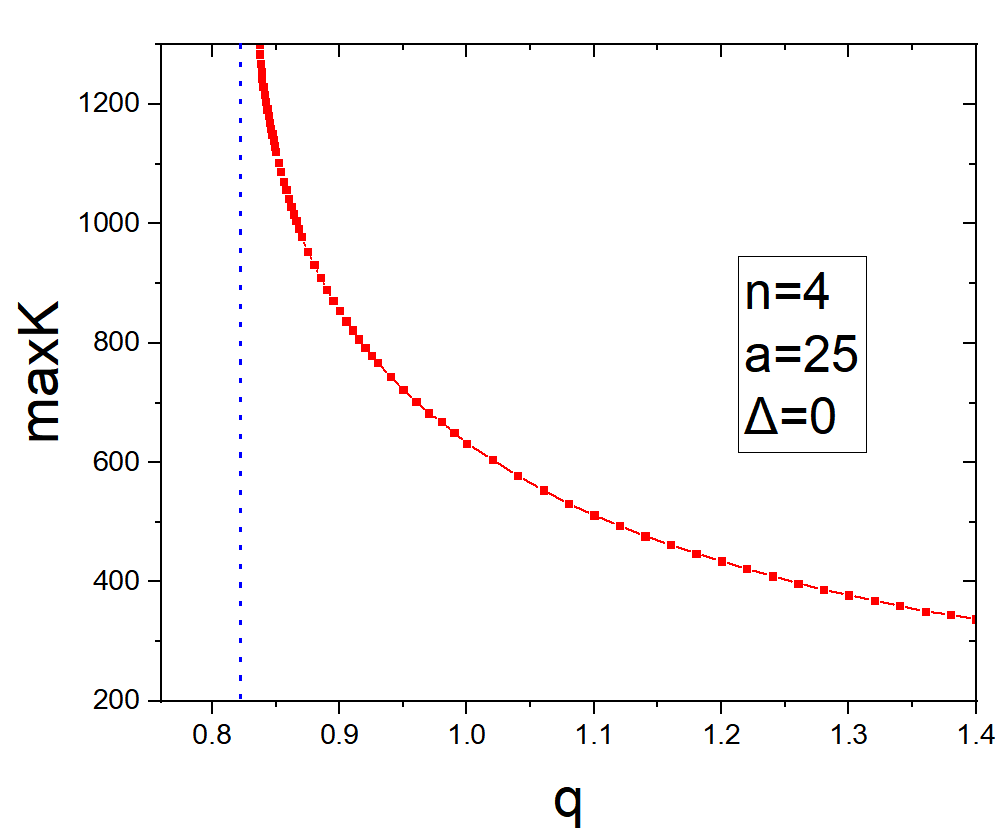}
\end{center}
\caption{Maximum of the Kretschmann scalar $K$ as a function of charge $q$, for $a=25$, $n=4$ and $\Delta=0$. The blue dashed line marks the minimal charge needed to save cosmic censorship.}
\label{fig:maxka}
\end{figure}
\begin{figure}[h!]
\begin{center}
\includegraphics[width=.4\textheight, angle =0]{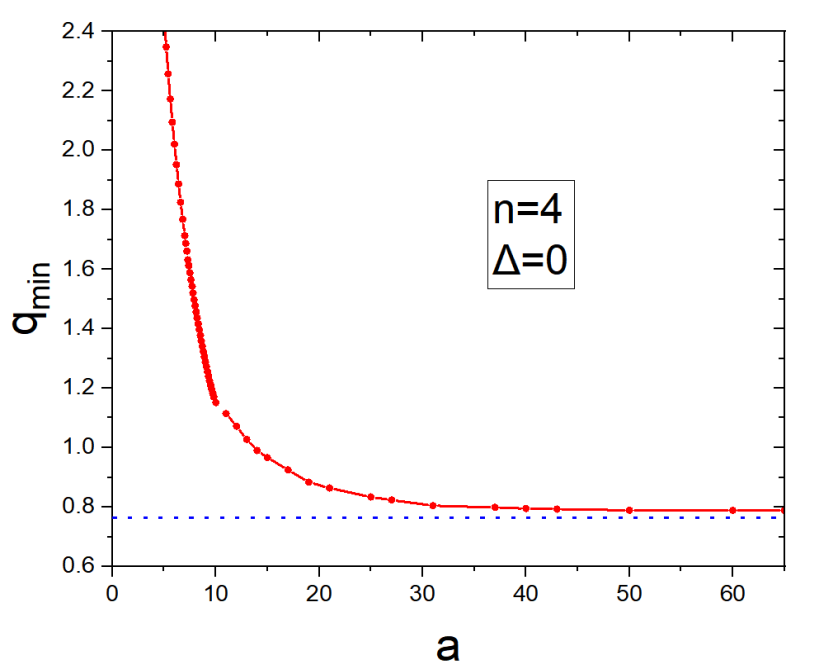}
\end{center}
\caption{\label{fig:aqmin}Phase diagram plotted for $\Delta=0$ and $n=4$. Minimal charge $q_{min}$ to preserve cosmic censorship approaches a constant $q_{W}$ for sufficiently large amplitudes $a$.}
\end{figure}

One natural problem is whether cosmic censorship can still be violated even though the $\phi$ component has been included, as we decrease the charge of  the $\phi$ component, just like the case of the charged scalar field discussed in \cite{Dias:2015nua}. To answer this we start with a smooth solution with $w(x, y)\neq0$, and lower the charge $q$ at a fixed amplitude $a=25$, until the solution becomes singular.
It turns out that there is indeed a minimal charge below which the spacetime curvature grows without bound. As shown in Fig.\ref{fig:maxka}, we plot the maximum value of the Kretschmann scalar $K$ as a function of $q$ for $n=4$, $a=25$ and $\Delta=0$. The vertical blue dotted line marks $q_{min}=0.825$, which is the minimal charge required to preserve cosmic censorship. Once the charge $q$ slightly below this minimal charge $q_{min}$, the maximum of Kretschmann scalar $K$ grows without bound, indicating the solution with $w(x,y)\neq0$ becomes singular.
We therefore argue that, even though solutions with the nonzero $\phi$ component indeed exist for large amplitude $a=25>a_{max}$, weak cosmic censorship conjecture might still be violated when the charge $q$ is sufficiently small. It is to say that, to preserve cosmic censorship the charge of the $\phi$ component has to be larger than the minimum value $q_{min}$.

We want to know if for any value of the amplitude $a>a_{max}$, there is always a minimal charge $q_{min}$ below which cosmic censorship is violated. To answer this we have repeated the calculation for the minimal charge for several different values of amplitude $a>a_{max}$. The results are displayed in Fig.\ref{fig:aqmin}. In Fig.\ref{fig:aqmin} we plot the minimal charge $q_{min}$ as a function of the boundary amplitude $a$ for $n=4$ and $\Delta=0$. There are two disconnected curves of the minimal charge, one of which just represents the minimal charge required to preserve cosmic censorship for $a> a_{max}$, and the other of which for $a< a_{max}$, denotes the onset of condensing the $\phi$ component. Above these two curves the smooth solutions with $w(x, y)\neq0$ always exist.

Unlike the results for the charged scalar field, the minimal charge of the $\phi$ component $q_{min}$ is a decreasing function of the boundary amplitude $a$, and eventually appears to approach a lowest constant value $q_W=0.78$, which is marked by the vertical blue line.
 Notice that with the restriction of calculation resource, we only display a part of the curve of the minimal charge for $a\leq60$. But it is clear that for large amplitudes $a$ the minimal charge $q_{min}$ required to save cosmic censorship only slightly above $q_W$, and obviously the difference $q_{min}-q_W$ always decreases as amplitude $a$ increases. 
 We therefore expect that as the amplitude $a$ increases, the minimal charge $q_{min}$ required to preserve cosmic censorship approaches the minimum value $q_W=0.78$. This means that including the $\phi$ component with charge larger than the minimum value $q_W$, when the boundary amplitude $a$ is sufficiently large, the original counterexamples in Einstein-Maxwell theory can be removed and the weak cosmic censorship conjecture is preserved.

\section{Conclusions}\label{sec4}
In this work, we construct the stationary solutions of the Einstein-Yang-Mills theory in four dimensional asymptotically anti-de Sitter spacetime. When the $SU(2)$ gauge field has only a nonzero $t$ component, we construct a class of counterexamples to the weak cosmic censorship conjecture in Einstein-Maxwell theory. However, including a nonzero $\phi$ component of $SU(2)$ gauge field with sufficiently large charge, the original counterexamples can be removed and cosmic censorship is preserved. Notice that the minimal charge $q_{min}$ required to preserve cosmic censorship decreases as amplitude $a$ increases and eventually approaches a constant value $q_W=0.78$. We thus argue that there is a minimum value $q_W$, when the charge of the $\phi$ component is larger than this minimum, and boundary amplitude $a$ is sufficiently large, the original counterexamples can be removed and the cosmic censorship can be saved.

However, unlike previous study of the relation between the scalar field charge and cosmic censorship, the study of the weak cosmic censorship conjecture with $SU(2)$ gauge field  shows a different point. When the amplitude is not large enough, even though the charge of the $\phi$ component is larger than the minimum value $q_W$, cosmic censorship can still be violated.
 This is because in Einstein-Yang-Mills theory the minimal charge $q_{min}$ of the $\phi$ component required to preserve cosmic censorship decreases as the amplitude $a$ increases, which is different from the case of a charged scalar field. 

In the further study we want to further explore the reason why for small amplitude $a$, the weak cosmic censorship can still be violated for $q>q_W$. There are two comments about this problem.
Firstly, in this work, we take the form of the Yang-Mills ansatz to have both the nonzero $t$ and $\phi$ component. However, since the vector potential of the Yang-Mills field is a matrix-valued function, in the future work, we can consider other nonzero components and use different ansatz of the Yang-Mills field. It is possible that there is a more appreciate form of the Yang-Mills field ansatz, such that assuming the charge larger than a minimum the cosmic censorship can be saved for all values of amplitude, rather than only for sufficiently large amplitudes.
Secondly, in this work we mostly focus on the Einstein-Yang-Mills solutions. In the future study, we can construct the stationary solutions of the Einstein-Maxwell-vector theory, to study the relation between the vector charge and cosmic censorship. We will discuss if there still is a minimum value such that cosmic censorship can be saved by assuming a vector field with charge larger than this minimum.

\section*{Acknowledgement}
We thank Yan-Bo Zeng and Shi-Xian Sun for helpful discussions.
This work is supported by National Key Research and Development Program of China (Grant No. 2020YFC2201503) and  the National Natural Science Foundation of China (Grant No.~12047501). Parts of computations were performed on the shared memory system at institute of computational physics and complex systems in Lanzhou university.

%%%%%%%%%%%%%%%%%%%%%%%%%%%%%%%%%%%%%%%%%%%%%%%%
%%%%%%%%%%%%%%%%%%%%%%%%%%%%%%%%%%%%%%%%%%%%%%%%

\end{document}